\documentclass{sig-alternate-nocopyright}

\usepackage{booktabs} 
\usepackage{caption} 

\usepackage{times}
\usepackage{url,bm}
\usepackage{latexsym}

\usepackage{epsfig}
\usepackage{subfigure}
\usepackage{xcolor}
\usepackage{amsmath,amssymb,color}
\usepackage{amssymb,color}
\usepackage{array}
\usepackage{algorithm}
\usepackage[noend]{algorithmic}
\usepackage{multirow}
\usepackage{graphics,epsfig}

\usepackage{etoolbox}
\newtoggle{conf} 
\toggletrue{conf} 
\newtoggle{msrtr}
\togglefalse{msrtr}

\newcommand{\cut}[1]{}

\newfont{\mycrnotice}{ptmr8t at 7pt}
 \newfont{\myconfname}{ptmri8t at 7pt}
 \let\crnotice\mycrnotice
 \let\confname\myconfname

\begin{document}

\title{Bringing Salary Transparency to the World: Computing\\ Robust Compensation Insights via LinkedIn Salary}

\numberofauthors{1}
\author{\alignauthor 
Krishnaram Kenthapadi, \ Stuart Ambler, \ Liang Zhang, \ Deepak Agarwal\\
\affaddr{LinkedIn Corporation, USA}\\
\email{(kkenthapadi, sambler, lizhang, dagarwal)@linkedin.com}
}

\maketitle

\begin{abstract}
The recently launched LinkedIn Salary product has been designed with the goal of providing compensation insights to the world's professionals and thereby helping them optimize their earning potential. We describe the overall design and architecture of the statistical modeling system underlying this product. We focus on the unique data mining challenges while designing and implementing the system, and describe the modeling components such as Bayesian hierarchical smoothing that help to compute and present robust compensation insights to users. We report on extensive evaluation with nearly one year of de-identified compensation data collected from over one million LinkedIn users, thereby demonstrating the efficacy of the statistical models. We also highlight the lessons learned through the deployment of our system at LinkedIn.
 \end{abstract}

\section{Introduction}\label{sec:intro}
Online professional social networks such as LinkedIn have enabled job seekers to discover and assess career opportunities, and job providers to discover and assess potential candidates.
For most job seekers, salary (or broadly, compensation) is a crucial consideration in choosing a new job opportunity\footnote{More candidates (74\%) want to see salary compared to any other feature in a job posting, according to a survey of over 5000 job seekers in US and Canada~\cite{careerBuilderSurvey2016}. Job seekers value compensation the most when looking for new opportunities, according to a US survey of 2305 adults~\cite{jobSeekerNationStudy2016}.}. At the same time, job seekers face challenges in learning the compensation associated with different jobs, given the sensitive nature of compensation data and the dearth of reliable sources containing compensation data. The recently launched LinkedIn Salary product\footnote{\url{https://www.linkedin.com/salary}} has been designed with the goal of providing compensation insights to the world's professionals, and thereby help them make more informed career decisions.

With its structured information including the work experience, educational history, and skills associated with over 500 million users, LinkedIn is in a unique position to collect compensation data from its users at scale and provide rich, robust insights covering different aspects of compensation, while preserving user privacy. For instance, we can provide insights on the distribution of base salary, bonus, equity, and other types of compensation for a given profession, how they vary based on factors such as region, experience, education, company size, and industry, and which regions, industries, or companies pay the most.

In addition to helping job seekers understand their economic value in the marketplace, the compensation data has the potential to help us better understand the monetary dimensions of the Economic Graph~\cite{Wei12} (which includes companies, industries, regions, jobs, skills, educational institutions, etc.).

The availability of compensation insights along dimensions such as gender, ethnicity, and other demographic factors can lead to greater transparency, shedding light on the extent of compensation disparity, and thereby help stakeholders including employers, employees, and policy makers to take steps to address pay inequality.

Further, products such as LinkedIn Salary can improve efficiency in the labor marketplace by reducing asymmetry of compensation knowledge, and by serving as market-perfecting tools for workers and employers~\cite{Har16}. Such tools have the potential to help students make good career choices, taking expected compensation into account, and to encourage workers to learn skills needed for obtaining well paying jobs, thereby helping reduce the skills gap.

We describe the overall design and architecture of the modeling system underlying LinkedIn's Salary product. 
We focus on unique challenges such as the simultaneous need for user privacy, product coverage, and robust, reliable compensation insights, and describe how we addressed them using mechanisms such as outlier detection and Bayesian hierarchical smoothing. 
We perform extensive evaluation with nearly one year of de-identified compensation data collected from over one million LinkedIn users, thereby demonstrating the efficacy of the statistical models. We also highlight the lessons learned through the production deployment of our system.

\section{Problem Setting}\label{sec:problem}
\begin{figure}
\centering
\includegraphics[width=0.45\textwidth]{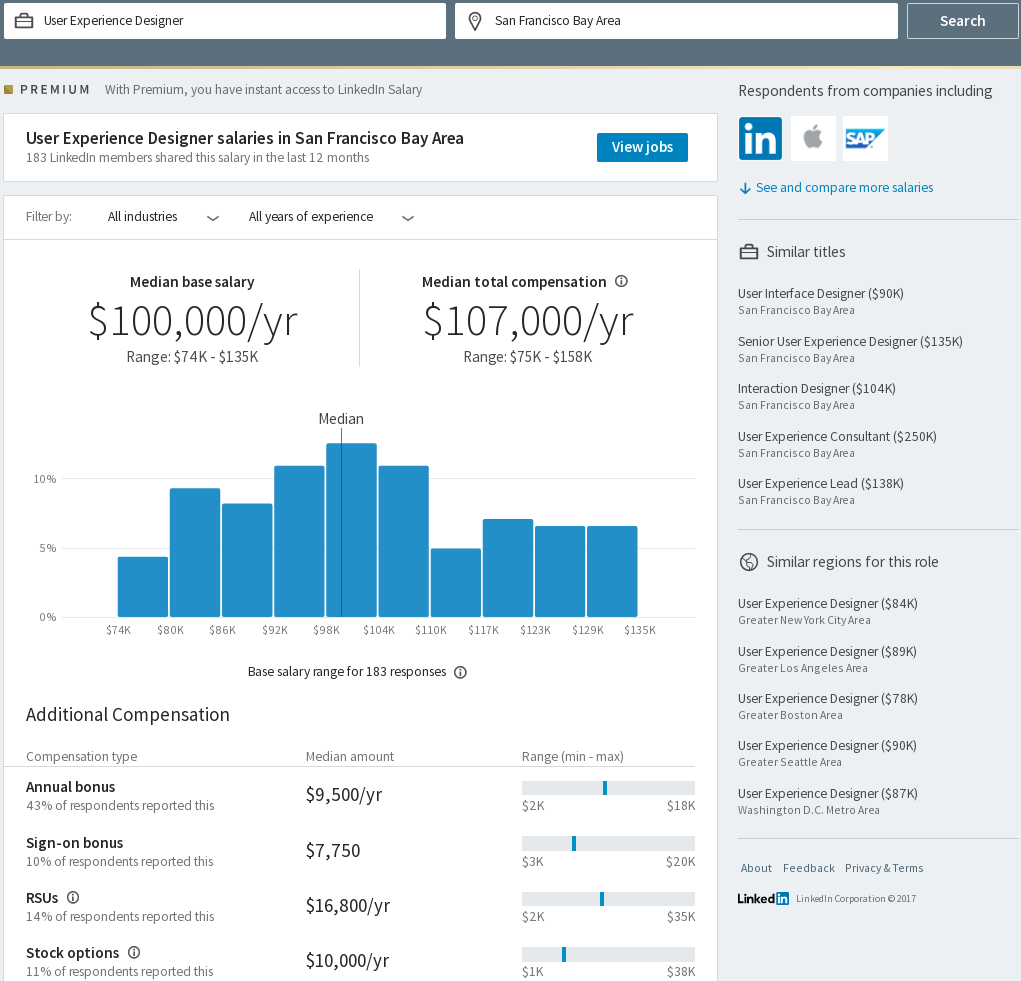}
\caption{LinkedIn Salary Insights Page}
\label{fig:salaryInsight}
\end{figure}

In the publicly launched LinkedIn Salary product~\cite{salaryEngBlogPost2016,salaryBlogPost2016}, users can explore compensation insights by searching for different titles and regions (Figure~\ref{fig:salaryInsight}). For a given title and region, we present the quantiles ($10$th and $90$th percentiles, median) and histograms for base salary, bonus, and other types of compensation. We also present more granular insights on how the pay varies based on factors such as region, experience, education, company size, and industry, and which regions, industries, or companies pay the most.

The compensation insights shown in the product are based on compensation data that we have been collecting from LinkedIn users. We designed a give-to-get model based data collection process as follows. First, cohorts (such as User Experience Designers in San Francisco Bay Area) with a sufficient number of LinkedIn users are selected. Within each cohort, emails are sent to a random subset of users, requesting them to submit their compensation data (in return for aggregated compensation insights later). Once we collect sufficient data, we get back to the responding users with the compensation insights, and also reach out to the remaining users in those cohorts, promising corresponding insights immediately upon submission of their compensation data.

Considering the sensitive nature of compensation data and the desire for preserving privacy of users, we designed our system such that there is protection against data breach, and any one individual's compensation data cannot be inferred by observing the outputs of the system. Our methodology for achieving these goals through a combination of techniques such as encryption, access control, de-identification, aggregation, and thresholding is described in~\cite{KCA17}. We next highlight the key data mining and machine learning challenges for the salary modeling system.

\subsection{Modeling Challenges}\label{sec:challenges}
{\em Modeling on de-identified data}: Due to the privacy requirements, the salary modeling system has access only to cohort level data containing de-identified compensation submissions (e.g., salaries for UX Designers in San Francisco Bay Area), limited to those cohorts having at least a minimum number of entries. Each {\em cohort} is defined by a combination of attributes such as title, country, region, company, and years of experience, and contains de-identified compensation entries obtained from individuals having the same values of those attributes. Within a cohort, each individual entry consists of values for different compensation types such as base salary, annual bonus, sign-on bonus, commission, annual monetary value of vested stocks, and tips, and is available without associated user name, id, or any attributes other than those that define the cohort. Consequently, our modeling choices are limited since we have access only to the de-identified data, and cannot, for instance, build prediction models that make use of more discriminating features not available due to de-identification.

{\em Evaluation}: In contrast to several other user-facing products such as movie and job recommendations, we face unique evaluation and data quality challenges. Users themselves may not have a good perception of the true compensation range, and hence it is not feasible to perform online A/B testing to compare the compensation insights generated by different models. Further, there are very few reliable and easily available ground truth datasets in the compensation domain, and even when available (e.g., BLS OES dataset~\cite{BlsOes08}), mapping such datasets to LinkedIn's taxonomy is inevitably noisy.

{\em Outlier detection}: As the quality of the insights depends on the quality of submitted data, detecting and pruning potential outlier entries is crucial. Such entries could arise due to either mistakes/mis\-understandings during submission, or intentional falsification (such as someone attempting to game the system).  We needed a solution to this problem that would work even during the early stages of data collection, when this problem was more challenging, and there may not be sufficient data across say, related cohorts.

{\em Robustness and stability}: While some cohorts may each have a large sample size, a large number of cohorts typically contain very few ($< 20$) data points each. Given the desire to have data for as many cohorts as possible, we need to ensure that the compensation insights are robust and stable even when there is data sparsity. That is, for such cohorts, the insights should be reliable, and not too sensitive to the addition of a new entry. A related challenge is whether we can reliably infer the insights for cohorts with no data at all.

Our problem can thus be stated as follows: {\em How do we design the salary modeling system to meet the immediate and future needs of LinkedIn Salary and other LinkedIn products? How do we compute robust, reliable compensation insights based on de-identified compensation data (for preserving privacy of users), while addressing the product requirements such as coverage?} We address these questions in \S\ref{sec:arch} and \S\ref{sec:model} respectively.
 \section{LinkedIn Salary Modeling System Design and Architecture}\label{sec:arch}
\begin{figure}[t]
\centering
\includegraphics[width=0.52\textwidth]{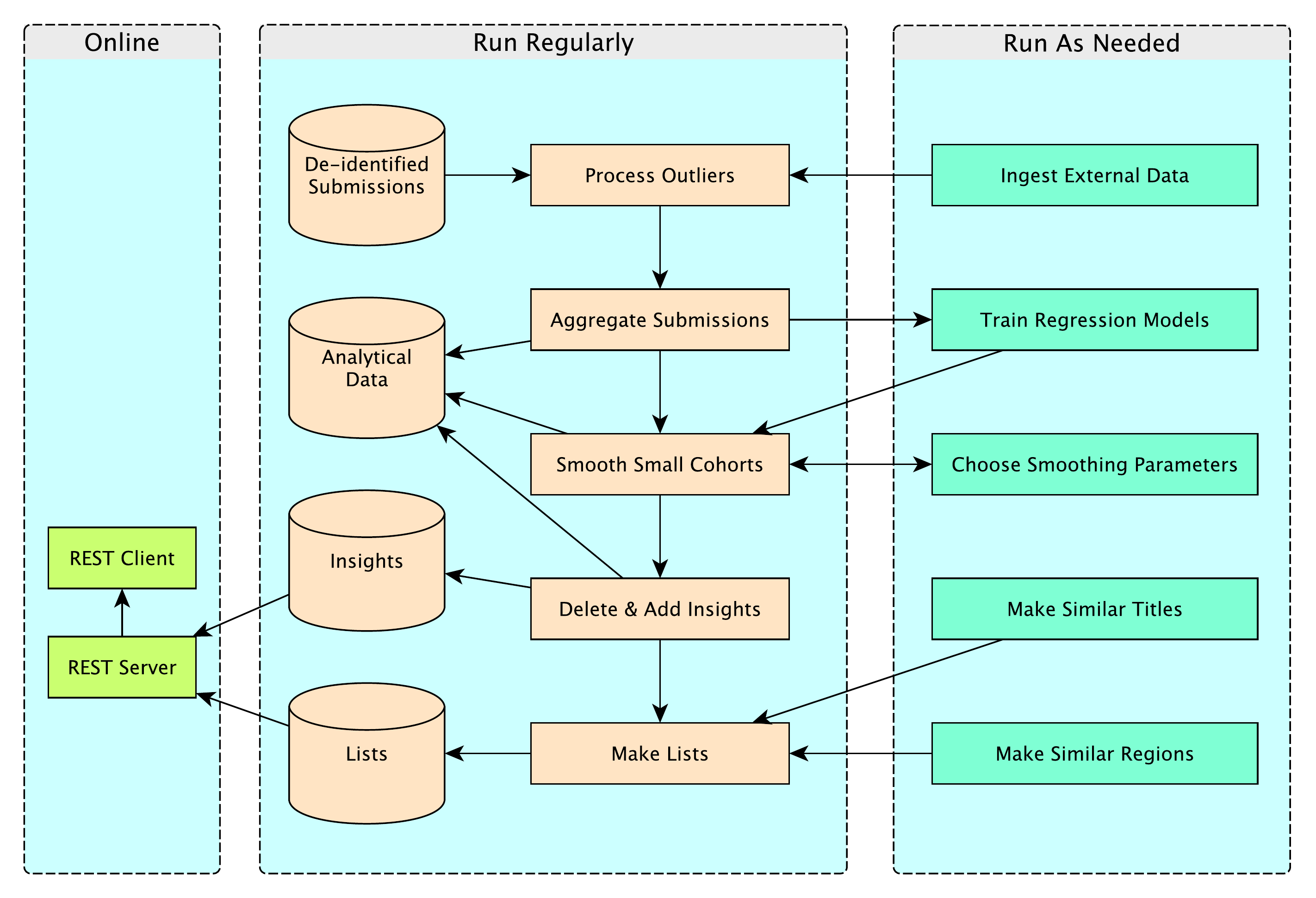}
\caption{Salary Modeling Online and Offline Architecture}
\label{fig:braavosrelevancearch}
\end{figure}

We describe the overall design and architecture of the salary modeling system deployed as part of the recently launched LinkedIn Salary product. Our system consists of an online component that uses a service oriented architecture for retrieving compensation insights corresponding to the query from the user facing product, and an offline component for processing de-identified compensation data and generating compensation insights. Figure~\ref{fig:braavosrelevancearch} presents the key components of our system, divided into three groups: {\em Online}, {\em Run Regularly}, and {\em Run As Needed}. Explanations of the text in the figure are in {\em italics} in the remainder of the section.

\subsection{Online System for Retrieving\\ Compensation Insights}\label{sec:online}

\subsubsection{LinkedIn Salary Platform}
The {\em REST Server} provides compensation insights on request by instances of {\em REST Client}.  The REST API~\cite{fielding2000architectural} allows retrieval of individual insights, or lists of them.
For each cohort, an insight includes, when data is available, the quantiles ($10$th and $90$th percentiles, median), and histograms for base salary, bonus, and other compensation types.
For robustness of the insights in the face of small numbers of submissions and changes as data is collected, we report quantiles such as $10$th and $90$th percentiles and median, rather than absolute range and mean.

\subsubsection{LinkedIn Salary Use Case}
For an eligible user, compensation insights are obtained via a {\em REST Client} from the {\em REST Server} implementing the Salary Platform REST API. 
These are then presented as part of LinkedIn Salary product. Based on the product and business needs, the eligibility can be defined in terms of criteria such as whether the user has submitted his/her compensation data within the last one year (give-to-get model), or whether the user has a premium membership.

Our Salary Platform has four service APIs to give the information needed for LinkedIn Salary insight page: (1) a ``criteria'' finder to obtain the core compensation insight for a cohort, (2) a ``facets'' finder to provide information on cohorts with insights for filters such as industry and years of experience, (3) a ``relatedInsights'' finder to obtain compensation insights for related titles, regions, etc., and (4) a ``topInsights'' finder to obtain compensation insights by top paying industries, companies, etc. These finders were carefully designed to be extensible as the product needs evolve over time. For instance, although we had originally designed the ``criteria'' finder to provide insights during the compensation collection stage, we were able to reuse and extend it for LinkedIn Salary and Job search use cases.

\subsubsection{Job Search Use Cases}
The compensation insights can be very valuable for enhancing other LinkedIn products. For example, we would like to present the compensation information as part of a webpage providing the description of a job posting, as part of the job search results page, and as a facet to filter jobs based on salary range.
Suppose the compensation insights are requested for a combination such as $\langle$title, company, region$\rangle$, e.g., $\langle$Software engineer, Slack Technologies, SF Bay Area$\rangle$ for presentation under a corresponding job posting webpage. We may not have sufficient data or confidence to return insights at this granularity. Hence, the Salary Platform {\em REST Server} generates possible generalizations such as  $\langle$Software engineer, Internet industry, SF Bay Area$\rangle$ or $\langle$Software engineer, SF Bay Area$\rangle$, and probes the Insights store for the corresponding precomputed insights. 

\subsection{Offline System for Computing\\ Compensation Insights}\label{sec:offline}

The insights are generated using an offline workflow, that consumes the {\em De-identified Submissions} data (corresponding to cohorts having at least a minimum number of entries) on HDFS, and then pushes the results to the {\em Insights} and {\em Lists} Voldemort key-value stores~\cite{sumbaly2012serving} for use by the {\em REST Server}.  {\em Analytical Data} is generated at several points for business analysis, modeling research, development, debugging, and testing.  We can also use previously generated compensation insights data for sanity checks on data quality, by checking differences in the size or number of cohorts, or in their insights.

This offline workflow consists of two groups, {\em Run Regularly} for the hadoop flow that runs more than once a day, and {\em Run As Needed} for hadoop and other flows that run, for instance, when externally supplied data changes.

\subsubsection{Hadoop Flow Run Regularly}
This consists of components that {\em Process Outliers} (see \S\ref{sec:outlier}), using several methods to detect questionable submissions and remove or modify them, {\em Aggregate Submissions} to obtain $10$th percentiles (``low end''), medians, $90$th percentiles (``high end''), and histograms of various kinds of compensation for each cohort, {\em Smooth Small Cohorts} (see \S\ref{sec:smoothing}) that uses Bayesian smoothing to help with some of the problems caused by cohorts with small numbers of submissions, {\em Delete \& Add Insights} to remove certain insights and add others such as from trusted reliable sources other than user submissions, and {\em Make Lists} to generate lists of insights or their keys.

\subsubsection{Flows Run As Needed}
These consist of components such as those that {\em Ingest External Data} (see \S\ref{subsubsec:externaloutlier}) to map external notions of title and region to their LinkedIn counterparts, {\em Train Regression Models} (see \S\ref{sec:regressionModel}) to train models used for smoothing and prediction, {\em Choose Smoothing Parameters} (see \S\ref{subsubsec:optimizesmoothingparameters}) to optimize tuning parameters for Baye\-sian smoothing, {\em Make Similar Titles} to create for each title, a list of similar ones, and its analog {\em Make Similar Regions}.

\section{Statistical Modeling for\\ Compensation Insights}\label{sec:model}

\subsection{Outlier Detection}\label{sec:outlier}
An important goal for the LinkedIn Salary product is accuracy, which is difficult to evaluate, since there are few reliable public datasets with comparable information.  We use user-submitted data as the main basis of reported results.  Even if submissions were completely accurate, there would still be selection bias; but accuracy of submissions cannot simply be assumed.  Mistakes and misunderstandings in submission entry occur, and falsification is possible.

As part of the salary modeling offline workflow, de-identified submissions are treated with three successive stages of outlier detection to reduce the impact of spurious data.  The first stage uses sanity-check limits such as the federal minimum wage as lower limit, and an ad-hoc upper bound (e.g., \$$2M$/year in US), for base salary. The second stage uses limits derived from US Bureau of Labor Statistics (BLS) Occupational Employment Statistics (OES) data~\cite{BlsOes08}, aggregated from federal government mail surveys. This dataset contains estimates of employment rates and wage percentiles for different occupations and regions in US.  The third stage is based on the internal characteristics of each cohort of submissions, using a traditional box-and-whisker method applied to the data remaining from the second stage.

\subsubsection{External Dataset based Outlier Detection}\label{subsubsec:externaloutlier}
We map the BLS OES compensation dataset into LinkedIn titles and regions as follows, for outlier detection for US base salary data.  There are about $840$ BLS occupations with SOC (Standard Occupational Classification) codes; an example is 13-2051, Financial Analysts.  To map them to the circa $25$K LinkedIn standardized (canonical) titles, first we expand them to O*NET alternate titles (look for alternate titles data dictionary at www.onetcenter.org).  Our example becomes $53$ alternate titles including Bond Analyst, Chartered Financial Analyst, Money Manager, Trust Officer, etc.  To these, we apply LinkedIn standardization software which maps an arbitrary title to a canonical title.  Bond Analyst maps to Analyst, and the other three to themselves.  In general, one BLS SOC code corresponds to more than one LinkedIn standardized title.  The mapping from BLS to LinkedIn regions is done using zipcodes.  In general, more than one BLS region corresponds to one LinkedIn region.  Thus, we have a many to many mapping.

For each LinkedIn (title, region) combination, we obtain all BLS (SOC code, region) rows that map to it, compute associated box-and-whisker compensation range limits, and aggregate these limits to derive one lower and one upper bound.

We obtained the limits for 6.5K LinkedIn standardized titles and about 300 LinkedIn US region codes, resulting in 1.5M $\langle$title, region$\rangle$ pairs.  Submissions with base salary outside these limits are excluded.

\subsubsection{Outlier Detection from User Submitted Data}\label{subsubsec:boxandwhisker}

Outlier detection based on user submitted data itself is done by a box-and-whisker method~\cite{engineeringStatisticsHandbook}\footnote{\url{http://www.itl.nist.gov/div898/handbook/prc/section1/prc16.htm}}, applied separately to each compensation type.

The box-and-whisker method is as follows.  For each compensation type for each cohort, we compute $Q1$ and $Q3$, the first and third quartiles respectively, the interquartile range, $IQR = Q3 - Q1$, then compute the lower limit as $Q1 - 1.5 \cdot IQR$, and the upper limit as $Q3 + 2 \cdot IQR$.
We chose (and tuned) the different factors $1.5$ and $2$ to reflect the typically skewed compensation distributions observed in our dataset.

Submissions with base salary outside the calculated range are excluded.  Other compensation type data are instead clipped to the limits.  We do not prune the entire entry, since the base salary is valid, and given that, do not want to remove outlying values of other compensation types, since that would have the effect of making them zero for the total compensation calculation.  We also exclude whole compensation types or even whole cohorts when the fraction of outliers is too large.

Note that this third stage is different in kind from the second stage, where the limits do not depend on the submissions received.
 
\subsection{Bayesian Hierarchical Smoothing}\label{sec:smoothing}
There is an inherent trade-off between the quality of compensation insights and the product coverage. The higher the threshold for the number of samples for the cohort, the more accurate the empirical estimates are. The lower the threshold for the number of samples, the larger the coverage. Since it is critical for the product to have a good coverage of insights for many different cohorts, obtaining accurate estimates of insights (e.g., percentiles) for cohorts with very few user input samples turns out to be a key challenge. Due to the sparsity of the data, empirical estimates of the 10th or 90th percentile, or even the median, are not reliable when a cohort contains very little data. For example, the empirical estimate of the 10th percentile of a cohort's compensation with only 10 points is the minimum of the 10, which is known to be a very inaccurate estimate.

We next describe a Bayesian hierarchical smoothing methodology to obtain accurate estimates of compensation insights for cohorts with very little data. Specifically, for cohorts with large enough sample sizes (i.e., number of samples greater than or equal to a threshold $h$, say 20), we consider it safe to use the empirical estimates for median and percentiles (see \S\ref{sec:lessons} for a discussion of the tradeoffs in choosing $h$). On the other hand, for cohorts with sample sizes less than $h$, we first assume that {\em the compensation follows a log-normal distribution}~\cite{pinkovskiy2009parametric} (see \S\ref{sec:lognormalvalidation} for validation of the assumption), then exploit the rich hierarchical structure amongst the cohorts, and ``borrow strength'' from the ancestral cohorts that have sufficient data to derive cohort estimates. For example, by successively relaxing the conditions, we can associate the cohort ``UX designers in SF Bay Area in Internet industry with $10+$ years of experience'' with larger cohorts such as ``UX designers in SF Bay Area in Internet industry'', ``UX designers in SF Bay Area with $10+$ years of experience'', ``UX designers in SF Bay Area'', and so forth, and pick the ``best'' ancestral cohort using statistical methods (\S\ref{sec:findParent}). After the best ancestor is selected, we treat the data collected from the ancestral cohort as the {\em prior} to apply a Bayesian smoothing methodology, and obtain the {\em posterior} of the parameters of the log-normal distribution for the cohort of interest (\S\ref{sec:smoothprior}). We also describe how to handle the ``root'' cohorts (a combination of job title and region in our product) using a prior from a regression model in \S\ref{sec:regressionModel}.

\subsubsection{Finding the Best Ancestral Cohort}\label{sec:findParent}
The set of all possible cohorts forms a natural hierarchy, where each cohort in the hierarchy can have multiple ancestors, as in the example of ``UX designers in SF Bay Area in Internet industry with $10+$ years of experience'' given above.
We describe our statistical approach to find the ``best'' ancestral cohort among all the ancestors of the cohort of interest in the hierarchy, where the ancestor that can ``best explain'' the observed entries in the cohort of interest statistically is considered the best. The ancestor can later be used to provide the prior information to estimate the posterior of parameters for the distribution of compensation insights for the cohort. We note that for ancestors with sample size greater than $h$, the empirical estimates of parameters are used to obtain the prior; otherwise, the posterior estimates of the parameters for the ancestor will be used as the prior of the cohort of interest.

We denote the cohort of interest as $y$, and the observed {\em log-transformed} compensation data that belong to $y$ as $y_i$, $i=1,\ldots,n$, with empirical average $\bar{y}$. 
Let $P=\{z^{(1)},\ldots, z^{(K)}\}$ be the set of $y$'s ancestral cohorts. Our objective is to pick the ``best'' ancestral cohort from $P$, based on how well the data in $y$ matches the data in the ancestor statistically.

Assume for cohort $z^{(j)}$, the compensation data follows a log-normal distribution, with mean $\mu_j$ and variance $\sigma^2_j$ after the log transformation. We can pick the best $z^{(J)}$ out of the K ancestors of $y$, by the following criteria (maximizing log-likelihood, or equivalently minimizing negative log-likelihood, assuming each $y_i$ to be independently drawn at random from the distribution of $z^{(j)}$), using the estimates of $\mu_j$ and  $\sigma^2_j$ for each $z^{(j)}$, $j=1,\ldots,K$:
\vspace{-0.1in}
\begin{equation}\label{eqn:bestParent}
J = \arg\min\limits_j \left( \frac{n}{2} \log(2\pi\sigma^2_j) + \frac{\sum\limits_{i=1}^n (y_i-\mu_j)^2}{2\sigma^2_j} \right)\\
\end{equation}
We then use the corresponding $\mu_J$ and $\sigma_J^2$ that provides the maximum of the log-likelihood as the prior for the smoothing of the compensation percentiles for cohort $y$ (see \S\ref{sec:smoothprior} for details). We also note that in Equation (\ref{eqn:bestParent}), if the number of samples for $z^{(j)}$ is greater than $h$, we compute $\mu_j$ and  $\sigma^2_j$ from the empirical estimates from its data directly. Otherwise,  $\mu_j$ and  $\sigma^2_j$ would be estimated by the posterior of cohort $z^{(j)}$ smoothed by the prior from $z^{(j)}$'s ancestors, again following \S\ref{sec:smoothprior}.

\subsubsection{Bayesian Hierarchical Smoothing}\label{sec:smoothprior}
For cohort $y$, suppose we pick cohort $z$ as the best ancestor following \S\ref{sec:findParent}. Denote $\mu$ and $\sigma^2$ as the estimated mean and variance for $z$. Also, denote number of samples in cohort $z$ as $m$. We can assume the following model for the data $y_i,$ $i=1,\ldots,n$, with $y_i$ being the log of each compensation sample in cohort $y$:
\begin{equation}
y_i \sim N(\nu, \tau^2), i=1,\ldots,n 
\end{equation}
where we assume $\nu$ and $\tau^2$ have the conjugate prior as follows:
\begin{equation}\label{eqn:nu}
\nu | \tau^2 \sim N(\mu, \frac{\tau^2}{n_0}), 
\end{equation}
\begin{equation}\label{eqn:tau}
\tilde{\tau}=1/\tau^2 \sim Gamma(\eta/\sigma^2, \eta), 
\end{equation}
where $n_0=m/\delta$, and $\delta$ and $\eta$ are hyper-parameters that can be tuned through cross-validation. We note that the prior mean of $\nu$ is the same as the best ancestor's mean $\mu$, and for the prior distribution of $\tilde{\tau}$, $1/\sigma^2$ is the prior mean (also from the best ancestor) and $1/(\eta\sigma^2)$ is the prior variance. 

Now we derive the posterior $p(\nu,\tilde{\tau}|y_1,\ldots,y_n)$. Let $\bar{y}=\sum\limits_i y_i/n$ be the empirical average of observations in $y$. First, the joint posterior,
\begin{equation}\label{eqn:jointPosterior}
p(\nu, \tilde{\tau}|y_1,\ldots,y_n) \propto p(y_1,\ldots,y_n|\nu,\tilde{\tau})p(\nu|\tilde{\tau})p(\tilde{\tau}). 
\end{equation}
The marginal posterior,
\begin{align}
p(\tilde{\tau}|y_1,\ldots,y_n) &\propto \int p(y_1,\ldots,y_n|\nu,\tilde{\tau})p(\nu|\tilde{\tau})p(\tilde{\tau}) d\nu  \notag \\
&\sim Gamma \left( \frac{n}{2}+\frac{\eta}{\sigma^2},\right. \notag \\
& \left. \eta+\frac{1}{2}\sum\limits_i (y_i-\bar{y})^2  + \frac{nn_0}{2(n+n_0)}(\bar{y}-\mu)^2 \right) .
\end{align}
Hence the posterior mean of $\tilde{\tau}$,
\begin{equation}
\hat{\tilde{\tau}}=\frac{\frac{n}{2}+\frac{\eta}{\sigma^2}}{\eta+\frac{1}{2}\sum\limits_i (y_i-\bar{y})^2 + \frac{nn_0}{2(n+n_0)}(\bar{y}-\mu)^2}. 
\end{equation}

For simplicity, we plug in $\hat{\tau}^2=1/\hat{\tilde{\tau}}$ as the estimate of $\tau^2$ for the rest of the calculation. Given $\tilde{\tau}$, the posterior of $\nu$ is
\begin{align}
p(\nu|\tilde{\tau},y_1,\ldots,y_n) &\propto p(y_1,\ldots,y_n|\nu,\tilde{\tau})p(\nu|\tilde{\tau})\notag\\
&\sim N(\frac{n}{n+n_0}\bar{y}+\frac{n_0}{n+n_0}\mu, \frac{1}{(n+n_0)\tilde{\tau}})
\end{align}

Therefore, for any new observation $y_{new}$,
\begin{equation}
E[y_{new}|y_1,\ldots,y_n] = E[\nu|\tilde{\tau},y_1,\ldots,y_n] = \frac{n}{n+n_0}\bar{y}+\frac{n_0}{n+n_0}\mu, 
\end{equation}
\begin{align}
Var[y_{new}|y_1,\ldots,y_n] &= E[Var[y_{new}|y_1,\ldots,y_n,\nu]] \notag \\
& + Var[E[y_{new}|y_1,\ldots,y_n,\nu]] \notag \\
= \hat{\tau}^2+ Var[\nu|y_1,\ldots,y_n] 
&= (1+\frac{1}{n+n_0}) \hat{\tau}^2 
\end{align}
To summarize, the median, 10th percentile, and the 90th percentile of the compensation for cohort $y$ can be estimated as follows:
\begin{itemize}
\item Input: Data for cohort y as $y_1,\ldots,y_n$; For best ancestral cohort $z$ of $y$, let $z$'s sample size be $m$, mean be $\mu$ and variance be $\sigma^2$. Compute the empirical average, $\bar{y}$ of $y_1,\ldots,y_n$.
\item Tuning parameters: $\delta$ and $\eta$, which can be optimized via cross-validation (see \S\ref{subsubsec:optimizesmoothingparameters}). 
\item Let $n_0=m/\delta$.
\item Estimate the posterior mean of $\tau^2$ as
\begin{equation}
\hat{\tau}^2=\frac{\eta+\frac{1}{2}\sum\limits_i (y_i-\bar{y})^2 + \frac{nn_0}{2(n+n_0)}(\bar{y}-\mu)^2}{\frac{n}{2}+\frac{\eta}{\sigma^2}}. \notag
\end{equation}
\item For a new observation added to the population of existing observations of $y$, denoted as $y_{new}$, the mean and variance of $y_{new}$ (post log-transformation) are
\begin{equation}
\hat{\mu} = E[y_{new}|y_1,\ldots,y_n]= \frac{n}{n+n_0}\bar{y}+\frac{n_0}{n+n_0}\mu, \notag
\end{equation}
\begin{equation}
\hat{\sigma}^2 = Var[y_{new}|y_1,\ldots,y_n]= (1+\frac{1}{n+n_0}) \hat{\tau}^2. \notag
\end{equation}
Since normal distribution is symmetric, the median of log(compensation) is set to $\hat{\mu}$, 10th percentile to $\hat{\mu} - 1.282 \cdot \hat{\sigma}$, and 90th percentile to $\hat{\mu} + 1.282 \cdot \hat{\sigma}$.
\item Obtain the final estimates by taking exponential transformation of these three quantities.
\end{itemize}

\subsubsection{Smoothing for Root Cohorts} \label{sec:regressionModel}
We note that there can be cases where the root cohorts in the hierarchy do not have enough samples to estimate percentiles empirically. For LinkedIn Salary product specifically, we consider a root cohort as a function of a job title and a geographical region, and model the compensation for these cohorts using a number of features. Simply using parent cohorts such as title only or region only as the prior might not be good enough, as there is a lot of heterogeneity for compensation of the same title for different regions (e.g., New York vs. Fresno), and that of the same region for different titles as well (e.g., Software engineers vs. Nurses). Hence, for these cohorts, we build a feature based regression model to serve as the prior for Bayesian smoothing.

Suppose there are $P$ root cohorts, with a combined total of $N$ samples. Denote the value of $i$th sample for cohort $p$, after applying log-transformation as $u_{ip}$, and the vector of all responses as $\bm{U}$. Let the feature vector for cohort $p$ be $\bm{x}_{p}$, which can include features such as title, country, region, skills related to the title, average compensation data for the region determined from external sources, and so forth, and the entire design matrix be $\bm{X}$. The regression model can then be expressed as
\begin{equation}
u_{ip} \sim N(\bm{x}_p'\bm{\beta},\gamma^2),
\end{equation}
where $\bm{\beta}$ has a Gaussian prior
\begin{equation}
\bm{\beta}\sim N(0,\lambda \bm{I}),
\end{equation}
and $\lambda$ is a L2 penalty tuning parameter for the regression model. The estimate of $\bm{\beta}$ is given by
\begin{equation}
\hat{\bm{\beta}} = (\bm{X}'\bm{X}+\lambda \bm{I})^{-1}\bm{X}'\bm{U},
\end{equation}
and the estimate of $\gamma^2$ is given by
\begin{equation}
\hat{\gamma}^2=\frac{\sum\limits_i\sum\limits_p (u_{ip}-\bm{x}_p'\hat{\bm{\beta}})^2}{N}.
\end{equation}

For any root cohort, to obtain its posterior of $\nu$ and $\tilde{\tau}$ following Equation (\ref{eqn:jointPosterior}), the parameters $\mu$ and $\sigma^2$ for the prior in Equations (\ref{eqn:nu}) and (\ref{eqn:tau}) can be specified as: $\mu=\bm{x}_p'\hat{\bm{\beta}}$, and $\sigma^2=\hat{\gamma}^2$. 
We set the ``sample size'' $m$ to be equal to the smoothing sample size threshold $h$, with the interpretation that a regression model prior is given the same weight as an ancestral cohort with $h$ samples.

\begin{table}
\tiny
\captionsetup{width=0.6\textwidth}
\caption{Fractional Change From Adding Spurious Data}
\centering
\begin{tabular}{l r r r r r r r }
\hline\hline
Percent Added Data & $5$ & $10$ & $15$ & $20$ & $25$ & $30$ & $35$ \\
\hline
Adding Minimum Wage Data & & & & & & & \\
\hline
Mean Percent of Added, Removed & 45 & 37 & 29 & 22 & 14 & 5 & 0 \\
Mean Percent of Original, Removed & 0 & 0 & 0 & 0 & 0 & 0 & 0 \\
Mean Percent Change $10$th Percentile & -5 & -23 & -42 & -47 & -52 & -59 & -63 \\
Mean Percent Change Median & -1 & -2 & -4 & -6 & -8 & -10 & -11 \\
Mean Percent Change $90$th Percentile & -1 & -1 & -2 & -2 & -3 & -4 & -4 \\
\hline
Adding Data Between $10$th and $90$th Percentiles& & & & & & & \\
\hline
Mean Percent of Added, Removed & 0 & 0 & 0 & 0 & 0 & 0 & 0 \\
Mean Percent of Original, Removed & 0 & 0 & 0 & 0 & 0 & 0 & 0 \\
Mean Percent Change $10$th Percentile & 1 & 2 & 2 & 3 & 3 & 4 & 4 \\
Mean Percent Change Median & 0 & 0 & 1 & 1 & 1 & 1 & 2 \\
Mean Percent Change $90$th Percentile & -1 & -1 & -2 & -2 & -2 & -2 & -3 \\
\hline
Adding \$$2$ Million Data & & & & & & & \\
\hline
Mean Percent of Added, Removed & 100 & 100 & 100 & 100 & 100 & 92 & 0 \\
Mean Percent of Original, Removed & 0 & 0 & 0 & 0 & 0 & 0 & 0 \\
Mean Percent Change $10$th Percentile & 0 & 0 & 0 & 0 & 0 & 0 & 6 \\
Mean Percent Change Median & 0 & 0 & 0 & 0 & 0 & 1 & 13 \\
Mean Percent Change $90$th Percentile & 0 & 0 & 0 & 0 & 0 & 180 & 2300 \\
\hline
\multicolumn{8}{l}{Notes: percents rounded to integers with two significant digits.  Means are over all test cohorts.}
\end{tabular}
\label{tab:outlierdetectionaddedlowdata}
\end{table}

\section{Experiments}\label{sec:exp}
We next present an evaluation of our system for computing compensation insights as part of LinkedIn Salary, focusing on the statistical models. We study the impact of the outlier detection methods and evaluate the effectiveness of the regression model and Bayesian statistical smoothing. We also investigate the performance of the statistical smoothing techniques for different segments and under different parameter settings.

\subsection{Experimental Setup}\label{sec:expsetup}
As described earlier, we implemented our statistical models as part of the offline system for computing compensation insights, and deployed in production as part of LinkedIn Salary product. Our offline system for computing compensation insights is implemented in a distributed computing environment using Hadoop.

We performed our experiments over nearly one year of de-identified compensation data collected from more than one million LinkedIn users, across three countries (USA, Canada, and UK). Due to the privacy requirements, we have access only to cohort level data containing de-identified compensation submissions (e.g., salaries for Software Engineers in San Francisco Bay Area), limited to those cohorts having at least a minimum number of entries. Each cohort consists of a set of de-identified compensation data corresponding to individuals, and each individual entry consists of values for different compensation types.  There are approximately $54$K title-country-region cohorts, and $24$K title-country-region-company-cohorts, for which we have insights.

As discussed in \S\ref{sec:challenges}, we cannot compare the performance of different models using online A/B testing, since users may not necessarily have a good perception of the true compensation range. We also face additional evaluation challenges since there are very few reliable and easily available ground truth datasets in the compensation domain (e.g., BLS OES dataset), and it is challenging to map such datasets to LinkedIn's taxonomy due to the noise introduced.

\subsection{Studying Impact of Spurious Added Data}\label{sec:evaloutlier}

Because of the lack of reliable ground truth data, our focus is on studying the impact of adding spurious data, then winnowing it by the box-and-whisker method (\S\ref{subsubsec:boxandwhisker}).  We take the actual compensation data submitted, after it has gone through all the outlier detection stages, and consider it as valid.  This constitutes about $80$\% of the submissions in cohorts.  We limit this data to title-country-region cohorts, to US data of yearly base salary, and to cohorts with at least $20$ valid entries, the last so that statistical smoothing does not interfere with interpretation.  This leaves about $10$K cohorts for the study.

For each cohort, we record selected quantiles ($10$th percentile, median, and $90$th percentile), then perturb the cohort in three ways by addition of spurious, synthetic salary entries numbering certain fractions ($5$\% through $35$\% in steps of $5$\%) of the original numbers of entries.  The three methods of perturbation are:
\begin{itemize}
  \setlength{\itemsep}{1pt}
  \setlength{\parskip}{0pt}
  \setlength{\parsep}{0pt}
  \item Add the spurious data at the federal minimum wage, \$$15080$.
  \item Add the spurious data at \$$2$ million.
  \item Add the spurious data uniformly at random between the $10$th and $90$th percentiles, the low and high ends.
\end{itemize}
For each method, we examine the following, averaged over all the cohorts used in the test.
\begin{itemize}
  \setlength{\itemsep}{1pt}
  \setlength{\parskip}{0pt}
  \setlength{\parsep}{0pt}
  \item The fraction of the added entries that are removed as outliers; a perfect score is 1.
  \item The fraction of the original entries that are removed as outliers; a perfect score is 0.
  \item The fractional changes in the selected quantiles.  This is the most important of these measures in practice, because these are what we report to users.
\end{itemize}
For all the data added between low and high ends, less than $1$\% of the added or original entries were removed. 
The reported quantiles changed by less than 5\%.

For added high entries, about $100$\% were removed as outliers until the amount of the added data was sufficiently large: at $30$\%, $92$\% were removed, and at $35$\%, none.  Less than $1$\% of the original data were removed as outliers.  The selected quantiles changed by less than $1$\% until at $30$\% added, the median increased by 1\% and the high end increased by $180$\%, and at $35$\%, the low end increased by $6$\%, median by $13$\%, and high end by $2300$\%.

For added low entries, note that the minimum wage is in general closer to the low end of the original data, than \$$2$ million is to the high end.  One might therefore suspect that outlier detection by the box-and-whisker method of these added entries, would be less successful than for added high entries.  Little of the original data were eliminated as outliers; see Table~\ref{tab:outlierdetectionaddedlowdata} for other results.

Overall, the box-and-whisker outlier detection, aside from very large amounts of spurious data or when the low end changed from the addition of more than $5$\% spurious minimum wage data, did at least reasonably and often quite well.
 
\begin{figure}
\centering
\includegraphics[width=0.35\textwidth]{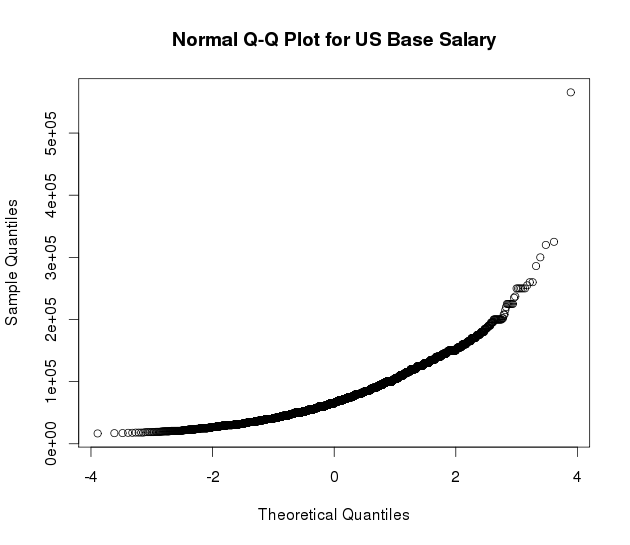}
\includegraphics[width=0.35\textwidth]{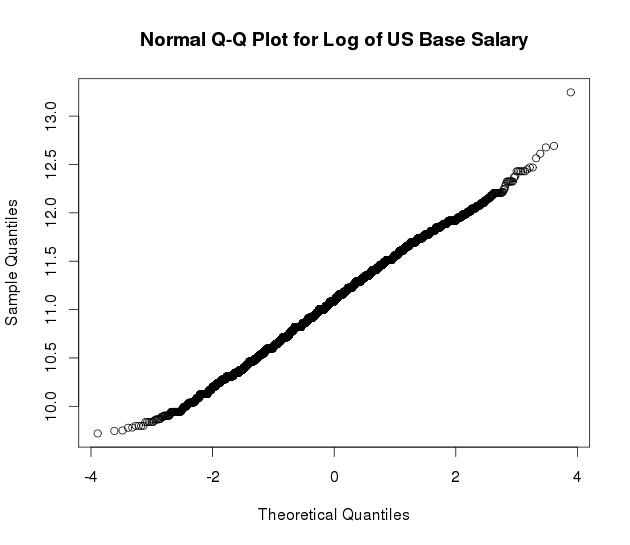}
\caption{Validation of Log-normal Distribution Assumption using Q-Q Plots. In both plots, X-axis corresponds to normal theoretical quantiles. Y-axis corresponds to the data quantiles for US base salaries in original and logarithmic scale respectively.}
\label{fig:qqplot}
\end{figure}

\subsection{Evaluating Regression Model and\\ Bayesian Statistical Smoothing}\label{sec:smoothingevaluation}
We first selected the set of base salaries reported by LinkedIn users in US, resulting in about 800K entries. We limited to only title-country-region cohorts so that each user submission is reflected exactly once. We used this dataset for validating the log-normal distribution assumption and for training and evaluating the regression model.

\subsubsection{Validation of Log-Normal Assumption}\label{sec:lognormalvalidation}
We sampled 10K entries at random from the above dataset, and generated Q-Q plots in both original and logarithmic scale. From Figure~\ref{fig:qqplot}, we can visually observe that the logarithmic scale is a better fit (closer to a straight line), suggesting that the data on the whole can be fit to a log-normal distribution. Hence, for training the regression model as well as applying statistical smoothing, we use the compensation data in the logarithmic scale. However, we remark that this assumption is an approximation since the observed salaries for an individual cohort may not follow a log-normal distribution.

\subsubsection{Evaluating Regression Model for Root Cohorts}
We trained and validated log-linear regression models for estimating compensation for each title-country-region cohort. A total of 24 models were trained corresponding to three countries and eight compensation types. We performed evaluation for all 24 combinations, but present results only for base and total compensation models in US.
For the (log) base salary model, we observed that the mean cross-validation error (0.085) is comparable to the mean empirical variance within each cohort (0.080). The corresponding numbers for the total compensation model were 0.097 and 0.092 respectively. Since we used only cohort-level features and not features that could discriminate samples within a cohort (due to de-identification), we do not expect the mean error to be less than the mean variance within each cohort. Thus, these results suggest that the trained models are performing well.

We next computed the coefficient of determination, $R^2$ for the (log) base salary model in two different ways. First, we computed over the entire dataset, where each row represents the de-identified (log) base salary reported by some user. There are as many rows for each title-country-region cohort as the number of users that provided salary for that combination. We computed $R^2 = 1 - \frac{\text{the residual sum of squares}}{\text{the total sum of squares}} = 0.58$. The same value (after rounding) was observed for the total compensation as well.
Due to de-identification however, intuitively, our model attempts to predict the mean log base salary for a cohort, but not the base salary for each user.
Hence, we also computed a variant of $R^2$ over the aggregated dataset, with one row per title-country-region cohort. The empirical mean log base salary for each cohort is treated as the observed value, and the predicted log base salary is treated as the predicted value. The coefficient computed in this manner was significantly higher: 0.88 for both base and total compensation. We observed similar results for all 24 models.

\subsubsection{Optimization of Statistical Smoothing\\ Parameters}\label{subsubsec:optimizesmoothingparameters}
We next describe how we chose the optimal values of the following parameters used as part of statistical smoothing:
(1) $\delta$, which is a discounting factor for the effect of the ancestral cohort, and
(2) $\eta$, which is a parameter used in the prior Gamma distribution for choosing the variance. 
The key idea is to maximize the likelihood of observing a hold-out set, with respect to the posterior distribution computed with each choice of smoothing parameters. We partitioned the set $S$ of all observed entries, restricted to cohorts with small sample size ($< 20$), where we applied smoothing, into a training set, $S_{train}$ and a hold-out test set, $S_{test}$ of 10\% of the data. We performed this over all possible cohorts (title-country-region, title-country-region-company, etc.), randomly partitioning each cohort, ensuring that at least one entry is present in the hold-out set. For different choices of smoothing parameters, we computed the posterior distribution for each cohort based on the training set, and computed the (combined) likelihood of observing the entries in the hold-out set (assuming independence across entries, which is an approximation). We selected the parameters that maximized this likelihood. Algorithm~\ref{alg:optimizesmoothingparameters} provides a formal description.

\begin{algorithm}
\caption{Optimizing smoothing parameters}\label{alg:optimizesmoothingparameters}
\begin{algorithmic}[1]
  \STATE \textbf{Input}: The set, $S$ of all observed entries, partioned into the training set, $S_{train}$ and the hold-out set, $S_{test}$; The candidate set, $\Delta$ of choices for $\delta$; The candidate set, $E$ of choices for $\eta$.
  \STATE \textbf{Output}: Optimal parameters, $\delta^*, \eta^*$.
  \FORALL{$\delta \in \Delta$}
    \FORALL{$\eta \in E$}
      \STATE Compute smoothed posterior log-normal distribution, $D_{posterior}(c)$ for each cohort, $c$ based on only $S_{train}$.
      \STATE $LL[\delta, \eta] := \sum_{s \in S_{test}} log(p_{D_{posterior}(c(s))}(s))$, where $c(s)$ denotes the cohort containing entry $s$.
    \ENDFOR
  \ENDFOR
  \STATE Output $(\delta^*, \eta^*) = \arg \max_{\delta \in \Delta, \eta \in E} LL[\delta, \eta]$.
\end{algorithmic}
\end{algorithm}

We performed grid search with $\Delta = [1, 1000]$ and $E = \{0.01 \cdot 2^r | r \in [0, 11] \}$, and computed the optimal parameters overall as well as for different segments such as cohorts containing a company, cohorts containing an industry, and so on. In all cases, the likelihood was maximized well inside this range (that is, not at the extremes for either $\delta$ or $\eta$). 
We observed that ($\delta^* = 5, \eta^* = 0.32$) was optimal overall, while for cohorts containing a company, ($\delta^* = 250, \eta^* = 0.04$), implying larger discount for ancestral cohort. This observation agrees with our intuition that the salaries at say, title-country-region-company level could deviate significantly from the corresponding title-country-region level (which is likely to be selected as the best ancestral cohort), and hence we would like to give relatively lesser weight to the ancestral cohort in such cases.

\subsubsection{Evaluating Statistical Smoothing:\\ Goodness-of-fit Analysis}\label{sec:goodnessoffit}
In addition to parameter optimization, Algorithm~\ref{alg:optimizesmoothingparameters} also helps us study the ``goodness-of-fit'' of the observed data in the hold-out set, with no smoothing vs. some extent of smoothing. Here, we compare the likelihood of observing each cohort in the hold-out set with respect to (1) a distribution derived just from the observed entries in the corresponding cohort in the training set, vs. (2) the smoothed posterior distribution that also takes ancestral cohort into account. As $\delta \rightarrow \infty$, the weight given to the ancestral cohort tends to zero, and hence the smoothed distribution converges to a (log-normal) distribution derived from just the observed entries. However, in \S\ref{subsubsec:optimizesmoothingparameters}, we observed that a finite value of $\delta$ ($= 5$ overall), rather than $\delta \rightarrow \infty$ (corresponding to the right extreme, $1000$ in our computation), leads to the maximum likelihood or the best goodness-of-fit. A similar observation holds even when limited to different segments such as cohorts containing a company. We conclude that combining ancestor and cohort data with statistical smoothing results in better goodness-of-fit for the observed data in the hold-out set, compared to using ancestor or cohort alone.

An intuitive explanation is that statistical smoothing provides better stability and robustness of insights. Inferring a distribution based on just the observed entries in the training set for a cohort and using it to fit the corresponding hold-out set is not as robust as using the smoothed distribution to fit the hold-out set, especially when the cohort contains very few entries. However, this evaluation approach intrinsically requires a distributional assumption for computing the likelihood, which can be thought of as a limitation since (1) any such assumption could be empirically validated on the whole (\S\ref{sec:lognormalvalidation}), but may not hold for an individual cohort, and (2) the log-normal distribution assumption is used as part of the smoothing methodology, and hence can be viewed as ``favoring'' it. Consequently, we decided to evaluate using a non-parametric approach that does not require any distributional assumption.

\subsubsection{Evaluating Statistical Smoothing:\\ Quantile Coverage Test}\label{sec:confinterval}
We next performed a quantile coverage test, wherein we measured what fraction of a hold-out set lies between two quantiles (e.g., 10th and 90th percentiles), computed based on the training set 
(1) empirically without smoothing, and (2) after applying smoothing.
For $0 < \alpha < \beta < 1$, an ideal quantile computation method should satisfy the following properties (averaged across all cohorts): (1) $\alpha$ fraction of the hold-out data points in a cohort should be lower than the corresponding $\alpha$-quantile, (2) $1-\beta$ fraction of the hold-out data points in a cohort should exceed the corresponding $\beta$-quantile, and (3) $\beta - \alpha$ fraction is in between, where the quantiles are computed for each cohort based on the training set entries. Since 10th and 90th percentiles are shown to the users in the product, we set $\alpha = 0.1, \beta = 0.9$.
We used the optimal parameters from \S\ref{subsubsec:optimizesmoothingparameters} for computing 10th and 90th percentiles using smoothing method.

We partitioned the set of observed entries (limited to cohorts with small sample size ($< 20$), where we applied smoothing) into a training set and a hold-out set as before. For each cohort, we computed 10th and 90th percentiles using the training set data points using each method (empirical vs. smoothed), and measured what fraction of the hold-out data points belong to this range. We then computed the mean fraction across all cohorts (and also for different segments of interest).

We observed that the fractions computed using smoothed percentiles are significantly better than those computed using empirical percentiles. The mean fraction of the hold-out data between 10th and 90th percentiles is 85\% with smoothing (close to the ideal of 80\%), but only 54\% with empirical approach. The remaining hold-out data is roughly evenly split below 10th percentile and above 90th percentile. 
We observed similar results for various segments such as cohorts containing a company, an industry, or a degree.

We also investigated the effect of the cohort size. Intuitively, we expect closer-to-ideal results for larger cohorts, but more deviation for very small cohorts. Comparing cohorts with at least 5 entries to those with just 3 or 4 entries, the above mean fraction deviates only slightly away from the 80\% ideal (83\% vs. 86\%), with smoothing applied. However, the deviation is significantly worse with empirical approach (71\% vs. 39\%), agreeing with our intuition that computing empirical percentiles based on very few entries is unreliable. 

While such a quantile coverage test by itself cannot imply that an approach is effective, the goodness-of-fit and coverage-based evaluations together establish that statistical smoothing leads to significantly better and robust compensation insights. 
Compensation domain knowledge experts also confirmed this conclusion.
By employing statistical smoothing, we were able to reduce the threshold used for displaying compensation insights in LinkedIn Salary product, thereby achieving significant increase in product coverage, while simultaneously preserving the quality of the insights.

\vspace{-0.1in}
\section{Lessons Learned in Practice}\label{sec:lessons}
We next present the challenges encountered and the lessons learned through the production deployment of both our salary modeling offline workflow and REST service as part of the overall LinkedIn Salary system for more than a year, initially during the compensation collection and later as part of the publicly launched product. We performed several iterative deployments of the offline workflow and the REST service, always maintaining a sandbox version (for testing) and a production version of our system. The service was deployed to multiple servers across different data centers to ensure better fault-tolerance and latency.

One of the greatest challenges has been the lack of good public ``ground truth'' datasets.  In the United States, the Internal Revenue Service has fairly complete salary data, but it is not public. The Bureau of Labor Statistics makes available aggregate data, but that brings with it the challenge of mapping to LinkedIn taxonomy.
Some state governments, for example California, make available government employee salaries, 
but government jobs differ from private-sector jobs.

We did simulations sampling smaller from larger cohorts to get an idea of how much variation in the reported quantiles might be expected if we decreased the sample size threshold. This analysis helped us understand the tradeoffs between increasing coverage and ensuring robust insights, and guided the product team decisions on whether to reduce the threshold. In fact, this analysis helped motivate the need for applying Bayesian smoothing, after which we were able to further reduce the threshold while retaining robustness of insights. The choice of smoothing threshold ($h = 20$) was determined by similar tradeoffs. While applying smoothing is desirable for even larger sized cohorts from the perspective of robustness, a practical limitation is that the smoothed histograms have to be computed based on a parametrized (log-normal) distribution, resulting in all smoothed histograms having identical shape (truncated log-normal distribution from 10th to 90th percentiles). As the empirical histogram was considered more valuable from the product perspective, we decided not to display any histogram for smoothed cohorts, and chose a relatively low threshold of 20 to ensure adequate coverage for cohorts with histograms.

  \section{Related Work}\label{sec:related}
{\em Salary Information Products}: There are several commercial services offering compensation information. For example, Glassdoor~\cite{GlaPr1} offers a comparable service, while PayScale~\cite{PayOne} collects individual salary submissions, offers free reports for detailed matches, and sells compensation information to companies.  The US Bureau of Labor Statistics~\cite{BlsOvr} publishes a variety of statistics on pay and benefits.

{\em Privacy}: Preserving user privacy is crucial when collecting compensation data.
We encountered unique challenges associated with privacy and security while designing our system. Our methodology for addressing these challenges through a combination of techniques such as encryption, access control, de-identification, aggregation, and thresholding is described in~\cite{KCA17}. Please refer this paper for a discussion of different privacy techniques (in contrast with our approach), and an empirically study of the tradeoffs between privacy and modeling needs.

{\em Survey Techniques}: There is extensive work on traditional statistical survey techniques~\cite{groves2011survey,jessen1978statistical}, as well on newer areas such as web survey methodology~\cite{callegaro2015web}. See~\cite{Ber05} for a survey of non-response bias challenges, and~\cite{bethlehem2010selection} for an overview of selection bias.

{\em Statistical Smoothing}: The idea of Bayesian hierarchical statistical smoothing originates from the smoothing of sparse events (e.g., CTR) in the context of computational advertising~\cite{agarwal2010estimating,zhang2009fast}, where a natural hierarchy for the combination of ads category and publisher category is used for an (ad, publisher) pair with very little data to borrow strength from its ancestor nodes. However, that is an entirely different context and the models used are hence different.

 \section{Conclusions and Future Work}\label{sec:conclusion}
We studied the problem of computing robust, reliable compensation insights based on de-identified compensation data collected from LinkedIn users. We presented the design and architecture of the modeling system underlying LinkedIn's Salary product, which was launched recently towards the goals of bringing greater transparency and helping professionals make more informed career decisions. We highlighted unique challenges such as modeling on de-identified data, lack of good evaluation datasets or measures, and the simultaneous need for user privacy, data quality, and product coverage, and described how we addressed them using mechanisms such as outlier detection and Bayesian hierarchical smoothing. We showed the effectiveness of our models through extensive experiments on de-identified data from over one million users. We also discussed the design decisions and tradeoffs while building our system, and the lessons learned from more than one year of production deployment.

The availability of compensation data, combined with other datasets, opens several research possibilities to better understand and improve the efficiency of career marketplace (as discussed in \S\ref{sec:intro}). There are also several directions to extend this work, which we are currently pursuing. We would like to improve quality of insights and product coverage via better data collection and processing, including inference of insights for cohorts with no data, improvement of the statistical smoothing methodology, better estimation of variance in the regression models, and creation of intermediary levels such as company clusters between companies and industries.
We also plan to improve outlier detection both at the user-level (using user profile and behavioral features, during submission), and at the cohort level (e.g., using the medcouple measure of skewness~\cite{HV07}). Another direction is to use other datasets (e.g., position transition graphs, salaries extracted from job postings) to detect/correct inconsistencies in the insights across cohorts.
Finally, mechanisms can be explored to quantify and address different types of biases such as sample selection bias and response bias. For example, models could be developed to predict response propensity based on user profile and behavioral attributes, which could then be used to compensate for response bias through techniques such as inverse probability weighting.

\section{Acknowledgments}
The authors would like to thank all other members of LinkedIn Salary team for their collaboration for deploying our system as part of the launched product, and
Stephanie Chou,
Ahsan Chudhary,
Tim Converse,
Tushar Dalvi,
Anthony Duerr,
David Freeman,
Joseph Florencio,
Souvik Ghosh,
David Hardtke,
Parul Jain,
Prateek Janardhan,
Santosh Kumar Kancha,
Ryan Sandler,
Cory Scott,
Ganesh Venkataraman,
Ya Xu,
and
Lu Zheng
for insightful feedback and discussions.
 
\bibliographystyle{abbrv}  
\bibliography{paper}

\begin{thebibliography}{10}

\bibitem{Ber05}
{\em Encyclopedia of Social Measurement}, volume~2, chapter Non-Response Bias.
\newblock Academic Press, 2005.

\bibitem{BlsOes08}
{\em {BLS Handbook of Methods}}, chapter 3, Occupational Employment Statistics.
\newblock U.S. Bureau of Labor Statistics, 2008.
\newblock http://www.bls.gov/opub/hom/pdf/homch3.pdf.

\bibitem{engineeringStatisticsHandbook}
{\em NIST/SEMATECH e-Handbook of Statistical Methods}.
\newblock National Institute of Standards and Technology, U.S. Department of
  Commerce, 2013.

\bibitem{careerBuilderSurvey2016}
How to rethink the candidate experience and make better hires.
\newblock {\em CareerBuilder's Candidate Behavior Study}, 2016.

\bibitem{jobSeekerNationStudy2016}
Job seeker nation study.
\newblock {\em Jobvite}, 2016.

\bibitem{GlaPr1}
Glassdoor introduces salary estimates in job listings, February 2017.
\newblock
  \url{https://www.glassdoor.com/press/glassdoor-introduces-salary-estimates-job-listings-reveals-unfilled-jobs-272-billion/}.

\bibitem{BlsOvr}
Overview of {BLS} statistics on pay and benefits, February 2017.
\newblock \url{https://www.bls.gov/bls/wages.htm}.

\bibitem{PayOne}
Payscale data \& methodology, February 2017.
\newblock
  \url{http://www.payscale.com/docs/default-source/pdf/data_one_pager.pdf}.

\bibitem{agarwal2010estimating}
D.~Agarwal, R.~Agrawal, R.~Khanna, and N.~Kota.
\newblock Estimating rates of rare events with multiple hierarchies through
  scalable log-linear models.
\newblock In {\em KDD}, 2010.

\bibitem{bethlehem2010selection}
J.~Bethlehem.
\newblock Selection bias in web surveys.
\newblock {\em International Statistical Review}, 78(2), 2010.

\bibitem{callegaro2015web}
M.~Callegaro, K.~L. Manfreda, and V.~Vehovar.
\newblock {\em Web survey methodology}.
\newblock Sage, 2015.

\bibitem{salaryEngBlogPost2016}
A.~Duerr and S.~K. Kancha.
\newblock Bringing salary transparency to the world.
\newblock {\em LinkedIn Engineering Blog}, 2016.
\newblock
  \url{https://engineering.linkedin.com/blog/2016/10/bringing-salary-transparency-to-the-world}.

\bibitem{fielding2000architectural}
R.~T. Fielding.
\newblock {\em Architectural styles and the design of network-based software
  architectures}.
\newblock PhD thesis, University of California, Irvine, 2000.

\bibitem{groves2011survey}
R.~M. Groves, F.~J. Fowler~Jr, M.~P. Couper, J.~M. Lepkowski, E.~Singer, and
  R.~Tourangeau.
\newblock {\em Survey methodology}.
\newblock John Wiley \& Sons, 2011.

\bibitem{Har16}
S.~Harris.
\newblock How to make the job market work like a supermarket.
\newblock {\em LinkedIn Pulse}, 2016.
\newblock
  \url{https://www.linkedin.com/pulse/how-make-job-market-work-like-supermarket-seth-harris}.

\bibitem{HV07}
M.~Hubert and E.~Vandervieren.
\newblock An adjusted boxplot for skewed distributions.
\newblock {\em Computational statistics \& data analysis}, 52(12), 2008.

\bibitem{jessen1978statistical}
R.~J. Jessen.
\newblock {\em Statistical survey techniques}.
\newblock John Wiley \& Sons, 1978.

\bibitem{KCA17}
K.~Kenthapadi, A.~Chudhary, and S.~Ambler.
\newblock {LinkedIn Salary}: A system for secure collection and presentation of
  structured compensation insights to job seekers.
\newblock In {\em IEEE PAC}, 2017.
\newblock {Available at \url{https://arxiv.org/abs/1705.06976}}.

\bibitem{pinkovskiy2009parametric}
M.~Pinkovskiy and X.~Sala-i Martin.
\newblock Parametric estimations of the world distribution of income, 2009.
\newblock Working Paper No. 15433, National Bureau of Economic Research.

\bibitem{salaryBlogPost2016}
R.~Sandler.
\newblock Introducing ``{LinkedIn Salary}'': Unlock your earning potential.
\newblock {\em LinkedIn Blog}, 2016.
\newblock
  \url{https://blog.linkedin.com/2016/11/02/introducing-linkedin-salary-unlock-your-earning-potential}.

\bibitem{sumbaly2012serving}
R.~Sumbaly, J.~Kreps, L.~Gao, A.~Feinberg, C.~Soman, and S.~Shah.
\newblock Serving large-scale batch computed data with project {Voldemort}.
\newblock In {\em FAST}, 2012.

\bibitem{Wei12}
J.~Weiner.
\newblock The future of {LinkedIn} and the {Economic Graph}.
\newblock {\em LinkedIn Pulse}, 2012.

\bibitem{zhang2009fast}
L.~Zhang and D.~Agarwal.
\newblock Fast computation of posterior mode in multi-level hierarchical
  models.
\newblock In {\em NIPS}, 2009.

\end{thebibliography}

\end{document}